\documentstyle[aps,preprint]{revtex}

\def\ZZ{{\mathchoice {\mathsf{Z \hspace{-0.45em} Z}} {\mathsf{Z 
        \hspace{-0.45em} Z}} {\mathsf{Z \hspace{-0.32em} Z}} 
    {\mathsf{Z \hspace{-0.23em} Z}}}}

\def\RR{{\mathchoice {\mathrm{I \hspace{-0.2em} R}} {\mathrm{I 
        \hspace{-0.2em} R}} {\mathrm{I \hspace{-0.14em} R}}
    {\mathrm{I \hspace{-0.14em} R}}}}

\begin{document}
\tighten
\title{\bf Scaling critical behavior of 
superconductors at zero magnetic field}

\author{C. de Calan and F. S. Nogueira}
\address{Centre de Physique Th\'eorique, Ecole Polytechnique, F-91128 
Palaiseau, FRANCE}

\date{Received \today}

\maketitle

\begin{abstract}
We consider the scaling behavior in the critical domain of superconductors 
at zero external magnetic field. The first part of the paper 
is concerned with the 
Ginzburg-Landau model in the zero magnetic field Meissner phase. We 
discuss the scaling behavior of the superfluid density and we give 
an alternative  
proof of Josephson's relation for a charged superfluid. This 
proof is obtained as a consequence of an exact renormalization 
group equation for the photon mass. We obtain Josephson's relation 
directly in the form $\rho_{s}\sim t^{\nu}$, that is, we do not need  
to assume that the hyperscaling relation holds. Next, we give an 
interpretation of a recent experiment performed in 
thin films of $YBa_{2}Cu_{3}O_{7-\delta}$. We argue that the 
measured mean field like behavior of the penetration depth exponent $\nu'$ 
is possibly associated with a non-trivial critical behavior 
and we predict the exponents $\nu=1$ and $\alpha=-1$ 
for the correlation lenght and specific heat, respectively.  
In the second part of the 
paper we discuss the scaling behavior in the continuum dual 
Ginzburg-Landau model. After reviewing lattice duality in 
the Ginzburg-Landau model, we discuss the  
continuum dual version by considering a family of scalings 
characterized by a parameter $\zeta$ introduced such that 
$m_{h,0}^2\sim t^{\zeta}$, where $m_{h,0}$ is the bare mass of the 
magnetic induction field. We discuss the difficulties in identifying the 
renormalized magnetic induction mass with the photon mass. 
We show that the only way to have a critical regime with 
$\nu'=\nu\approx 2/3$ is having $\zeta\approx 4/3$, that is, 
with $m_{h,0}$ having the scaling behavior of the renormalized photon 
mass.                 
\end{abstract}
\draft

\section{Introduction}

The study of the critical behavior of the high-temperature 
superconductors (HTSC) 
has been initiated long ago \cite{Lobb}. The discovery of this 
remarkable class of materials has opened a new perspective for the 
general theory of critical phenomena. In fact, the situation for the 
HTSC is much more favorable than for the conventional metallic 
low-temperature superconductors. The reason is that the size of the 
critical region in the HTSC is large enough to make its critical properties 
experimentally accessible using today techniques. In the tested temperature 
region it has been obtained 
that $-0.03<\alpha<0$ for the specific heat exponent, 
$\nu\approx 0.67$ for the correlation lenght exponent and the amplitude 
ratio $A^{+}/A^{-}\approx 1.065$ \cite{Salamon,Kamal}. 
These values are consistent with those 
found for $^4$He. This means that the critical region probed corresponds 
to an {\it uncharged} $3D$ $XY$ universality class \cite{Fisher}. In this 
situation the superfluid density, $\rho_{s}$, scales as 
$\rho_{s}\sim\lambda^{-2}$, where $\lambda$ is the penetration depth. 
The experimental result $\nu\approx 0.67$ follows from a direct 
measurement of $\lambda$. The measured value of the penetration 
depth exponent is $\nu'\approx 0.33$ and from Josephson's relation 
$\rho_{s}\sim\xi^{-1}$ the result $\nu\approx 0.67$ follows.   
Recent zero field experimental results obtained using 
very clean crystals of 
$YBa_{2}Cu_{3}O_{7-\delta}$ (YBCO) 
by Kamal {\it et al.} \cite{Kamal1},  
confirm early measurements of the penetration depth near $T_{c}$, 
giving $\nu'=0.33\pm 0.01$. These measurements estimate the critical region 
as being nearly $5$ K wide. While this seems to be true for 
three-dimensional crystals, it does 
not seem to be the case for thin films. Indeed, 
recent measurements of $\lambda$ in thin films of 
$YBa_{2}Cu_{3}O_{7-\delta}$ performed by Paget {\it et al.} 
\cite{Paget} display a 
critical regime where the $3D$ $XY$ behavior is absent and $\nu'=1/2$, that 
is, a mean field like behavior. The critical region reported by them is however 
only $0.5$ wide. The same result $\nu'=1/2$ has been obtained earlier 
for thin films by Lin {\it et al.} \cite{Lin}. However, the result of 
Lin {\it et al.} followed as a consequence of taking in account the finite 
size of the sample, otherwise the result $\nu'\approx 0.33$ of 
Kamal {\it et al.} \cite{Kamal} 
follows. We stress that none of these experiments were 
performed in the {\it charged} critical region.     
The true {\it charged} transition corresponds to a very  
small critical region and is presently inaccessible to the experimental 
probes. From duality arguments in the lattice model, it is obtained that 
the charged phase transition should correspond to an ``inverted'' 
$XY$ behavior \cite{Dasgupta}. 
For this situation it has been argued by 
Herbut and Tesanovi\'c \cite{Herbut} that $\rho_{s}\sim \lambda^{-1}$ 
which implies that $\nu'=\nu$. Recent numerical study in the lattice model 
\cite{Olsson} 
confirms this prediction and gives $\nu'=\nu\approx 0.67\approx 2/3$. 

The result $\nu\approx 0.67\approx 2/3$ 
has been obtained in recent years using 
perturbative renormalization group (RG) 
methods in {\it fixed} 
dimension $d=3$ \cite{Kiometzis,HerbutI,deCalan,Nogueira}. 
RG calculations performed in the 
seventies on the basis of the $\epsilon$-expansion leads 
to a flow where the charged infrared stable fixed point is absent if 
the number of components $n$ of the order parameter is less than $365.9$ 
\cite{HLM}. Since the 
physical case corresponds to $n=2$, we have that no second order 
phase transition is predicted in that calculation. The authors of 
ref. \cite{HLM} have predicted that a weak first order transition takes 
place. Further calculations using dimensional regularization 
confirmed this scenario even up to 2-loop order \cite{Hikami}. This 
picture seems to be appropriate for superconductors in the type I regime. 
However, for the type II regime this result is shown to be an 
artifact of the $\epsilon$-expansion \cite{Radz}. 
The $\epsilon$-expansion can be improved by doing a Pad\'e-Borel 
resummation \cite{Folk}. In this case a charged infrared stable fixed 
point is found and it is obtained that $\nu\approx 0.771$ \cite{Folk}. 
Non-perturbative 
calculations on the basis of the Wilson RG gives $\nu\approx 0.53$ 
by truncating the average action (the Legendre transform of the 
Wilsonian effective action) in $|\phi|^4$ and $\nu\approx 0.58$ with a 
truncation in $|\phi|^8$ \cite{Berg}. Thus, the 
perturbative RG in fixed dimension seems to give more interesting 
values of $\nu$ though it is less controlled. The non-perturbative 
calculations using Wilson RG of ref. \cite{Berg} are also performed in 
fixed dimension but it seems to be very difficult to make further 
improvements with respect to the different truncations. 

It is in general very difficult to find reliable approximations to study 
the critical domain of a charged transition. However, since the 
experimentally probed critical regime corresponds to a crossover near 
the neutral $XY$ universality class, theoretical studies of scaling 
behavior are often performed using a Ginzburg-Landau (GL) model 
coupled to an external 
magnetic field ${\bf H}$ 
and neglecting gauge field fluctuations \cite{Lawrie}. When the 
order parameter critical fluctuations are taken in account it is possible 
to study the different critical regimes in a phase diagram in the 
$H-T$ plane. The situation is particularly interesting for the HTSC 
where many new physical effects like  
vortex-lattice melting occurs \cite{Fisher,Brandt,Tesanovic}.

In this paper we will consider the scaling behavior in the 
{\it charged} critical 
domain in superconductors at the zero magnetic field Meissner phase. 
In the first part of the paper we will discuss the scaling behavior of the 
Ginzburg-Landau model. We use an exact RG equation for the 
photon mass to rederive the 
Herbut and Tesanovic result $\nu'=\nu$. Since our derivation does not 
use the Josephson relation \cite{Josephson}, we obtain as a consequence 
the Josephson relation for a charged superfluid directly in the 
form $\rho_{s}\sim t^{\nu}$, while in the original Josephson's paper this 
form follows only if it is assumed that hyperscaling holds since he has 
obtained actually that $\rho_{s}\sim t^{2\beta-\eta\nu}$. Next, we give an 
alternative 
interpretation of a recent experimental result of 
Paget {\it et al.} obtained using YBCO thin films \cite{Paget}.  

The second part of the paper is concerned with the continuum dual version of 
the GL model \cite{Kovner,Kiometzis,KiometzisI}. 
We review the lattice duality for the lattice GL model and 
the meaning of ``inverted'' $XY$ behavior \cite{Dasgupta}. 
Then we discuss the scaling behavior of the proposed continuum dual 
version of lattice duality. For this end, we consider a family of 
scalings $\Sigma_{\zeta}$ where the parameter $\zeta\geq 0$ 
is introduced through $m_{h,0}^2\sim t^{\zeta}$, with $m_{h,0}$ being 
the mass of the magnetic induction field. 
We obtain the 
scaling behavior for some relevant values of $\zeta$ by assuming that 
the renormalized counterpart of $m_{h,0}$ is the renormalized photon 
mass of the GL model in the Meissner phase. We show that the only way 
to obtain $\nu=\nu'\approx 2/3$ is using $\zeta\approx 4/3$.  
This means that $m_{h,0}$ has the same scaling behavior as the renormalized 
photon mass of the GL model. 
This analysis 
helps us to understand the difficulties in describing a correct 
``inverted'' $XY$ behavior with the continuum dual model.          
  
\section{Scaling in the GL model for $T<T_{c}$}
  
In the following we will assume the existence of an infrared 
stable fixed point. This should be true at least in the type II regime, as 
lattice results have convincingly shown \cite{Dasgupta,Olsson,Bart}. The 
existence of an infrared stable fixed point has also been established in 
recent years directly in the continuum model by using mainly RG techniques 
\cite{Herbut,Kiometzis,HerbutI,deCalan,Nogueira,Berg,Radz,Folk}. 
The bare action for the $d=3$ GL model in the zero external 
magnetic field Meissner phase is given by 

\begin{equation}
S({\bf A}_{0},\phi_{0};m_{0}^2,u_{0},e_{0})
=\int d^{3}x\left[\frac{1}{2}(\nabla\times{\bf A}_{0})^{2}
+|(\nabla-ie_{0}{\bf A}_{0})\phi_{0}|^{2}
-m_{0}^2|\phi_{0}|^{2}+\frac{u_{0}}{2}|\phi_{0}|^{4}\right],
\end{equation}
where the subindex $0$ denotes bare quantities and $m_{0}^2>0$. 
Here $m_{0}^2\sim t$, $t$ being the reduced temperature. 
The renormalized action,  
$S_{R}$, is defined in a standard way \cite{ZJ}, that is,

\begin{equation} 
S_{R}({\bf A},\phi;m^2,u,e)=
S(Z_{A}^{1/2}{\bf A},Z_{\phi}^{1/2}\phi;
Z^{(2)}_{\phi}Z_{\phi}^{-1}m^2,Z_{u}Z_{\phi}^{-2}u,
Z_{A}^{-1/2}e),
\end{equation}
where the quantities without the zeroes are renormalized 
and we have defined the corresponding renormalization constants \cite{Note1}. 
The renormalization constants above are the same as in the Coulomb phase, 
as dictated by the Ward identities \cite{Collins}. 
We assume that the correlation functions are evaluated in a renormalized 
gauge, the so called $R_{\alpha}$ gauge \cite{Collins}, and that the 
Coulomb gauge limit has been taken in the end. We cannot fix the gauge 
$\alpha=0$ from the very beginning because the unphysical fields 
constituting the $R_{\alpha}$ gauge would become massless, generating 
infrared divergences. The unitary gauge, on the other hand, does not have this 
problem since only physical fields are present. However, it is not 
renormalizable, even in $d=3$. 
 
In order 
to obtain the RG equations we must define a {\it scaling}, that is, the 
choice of the mass scale controling the flow while specifying   
the variables which are kept 
fixed in this process. Here we note that we have two 
dimensionful couplings, namely, $e^2$ and $u$, both having dimension of 
mass. In the Meissner phase the photon acquires a mass, 
$m_{A}$, which together with the mass, $m$, define two mass 
scales in the problem. An useful dimensionless parameter  
is the ratio between these 
two masses, the Ginzburg parameter, $\kappa=m/m_{A}=(u/e^2)^{1/2}$. Since 
the critical point corresponds to $m=0$, the renormalized mass $m$ is a 
good mass scale to control the flow. Thus, we will choose it as the 
fundamental mass scale. 
Let us define in this way the dimensionless 
couplings $\hat{u}\equiv um^{d-4}=Z_{\phi}^2 Z_{u}^{-1}u_{0}m^{d-4}$ and 
$\hat{e}^2\equiv e^2 m^{d-4}=Z_{A}e_{0}^2 m^{d-4}$. Note that although 
the GL model has been written in fixed dimension $d=3$, we have defined 
the dimensionless couplings in arbitrary dimension $2<d\leq 4$. 
In order to complete the definition of our scaling we 
must specify the parameters that will be kept fixed. The standard 
choice corresponds to differentiate the renormalized dimensionless couplings 
with respect to $\ln(m)$ keeping the bare couplings and 
the ultraviolet cutoff fixed. Note that in this 
process neither $m_{0}$ nor $m_{A,0}$ are kept fixed. 
We define the RG functions 

\begin{equation}
\eta_{A}=m\frac{\partial\ln Z_{A}}{\partial m}
\end{equation}, 

\begin{equation}
\eta_{\phi}=m\frac{\partial\ln Z_{\phi}}{\partial m}, 
\end{equation}

\begin{equation} 
\eta_{\phi}^{(2)}=m\frac{\partial\ln Z_{\phi}^{(2)}}{\partial m} 
\end{equation}
At the 
infrared stable fixed point $\eta_{A}$ and $\eta_{\phi}$ 
gives respectively the anomalous dimensions 
of ${\bf A}$ and $\phi$. The anomalous dimension of $|\phi|^2$ is 
given by the fixed point value of $\eta_{\phi}^{(2)}-\eta_{\phi}$.  
The fixed point values $\hat{e}^{2}_{*}$ and $\hat{u}_{*}$ 
are determined from the equations 
$\beta_{\hat{e}^2}\equiv m\partial\hat{e}^2/\partial m=0$ and 
$\beta_{\hat{u}}\equiv m\partial\hat{u}/\partial m=0$ and we take the 
solution corresponding to the infrared stable fixed point. The beta 
funcion $\beta_{\hat{e}^2}$ is given exactly by 

\begin{equation}
\label{be2}
\beta_{\hat{e}^2}=(\eta_{A}+d-4)\hat{e}^2.
\end{equation}     
An immediate consequence of the above equation is that a charged fixed 
point corresponds to $\eta_{A}^{*}=4-d$ \cite{Herbut,deCalan,Nogueira,Berg}, 
$\eta_{A}^{*}$ being the fixed point value of $\eta_{A}$. 
From Eq. (\ref{be2}) and the definition of 
$\kappa$ we have the following exact equations: 

\begin{equation}
\label{k}
m\frac{\partial\kappa^2}{\partial m}=\kappa^2\left(\frac{\beta_{\hat{u}}}{
\hat{u}}+4-d-\eta_{A}\right), 
\end{equation}

\begin{equation}
\label{mA}
m\frac{\partial m_{A}^2}{\partial m}=m_{A}^2\left(d-2+\eta_{A}-
\frac{\beta_{\hat{u}}}{\hat{u}}\right).
\end{equation}
From Eq. (\ref{mA}) we obtain easily that near the phase transition 
(that is, near the charged infrared stable fixed point) the photon mass 
scales as $m_{A}\sim m$. Since $m=\xi^{-1}$ and $m_{A}=\lambda^{-1}$, 
$\lambda$ being the 
penetration depth, we obtain $\lambda\sim\xi$ implying $\nu'=\nu$. This 
is a rederivation of the result of Herbut and Tesanovi\'c \cite{Herbut}. 
Note that in 
our derivation no use has been made of the Josephson relation. Since 
$m_{A}^2=e^2\rho_{s}$ and from Eq. (\ref{be2}) 
$e^2\sim m^{4-d}=\xi^{d-4}$, 
we obtain that $\rho_{s}\sim\xi^{2-d}\sim t^{\nu(d-2)}$. 
This constitutes a renormalization 
group proof of the Josephson's relation for the {\it charged} superfluid. 
In the original Josephson's paper this relation is proved for an 
uncharged superfluid and given in the form $\rho_{s}\sim t^{2\beta-\eta\nu}$ 
($\eta$ is the fixed point value of $\eta_{\phi}$). 
This form follows easily by noting that 
$\rho_{s}=<|\phi|^2>=Z_{\phi}^{-1}<|\phi_{0}|^2>$. Since 
near the critical point $Z_{\phi}\sim m^{\eta}\sim t^{\nu\eta}$ 
and defining $\beta$ through  
$<|\phi_{0}|^2>\sim t^{2\beta}$, it follows that  
$\rho_{s}\sim t^{2\beta-\nu\eta}$. Thus, in this last argument leading to   
Josephson's relation it does not matter if the superfluid is 
charged or not. Our derivation made directly for the superconductor 
implies therefore 

\begin{equation}
2\beta-\eta\nu=\nu(d-2).  
\end{equation}
As pointed out by Fisher {\it et al.} \cite{MEFisher} in the context 
of uncharged superfluids,  
the relation with the exponent $\nu(d-2)$ 
instead of $2\beta-\eta\nu$ holds only if hyperscaling holds, that is, 
$d\nu=2-\alpha$. Our argument shows that hyperscaling holds 
for a superconductor. 
For an uncharged superfluid the 
hyperscaling relation can be proved by 
using scaling RG arguments of the same type we used here \cite{ZJ}. 
Of course, this type of proof is not rigorous in the sense that it assumes 
that the continuum limit of the lattice statistical mechanical model 
exists. Only the inequality $d\nu\geq 2-\alpha$ can be rigorously proved  
\cite{Sokal}. A well known case where hyperscaling fails is mean field 
theory where $\nu=1/2$ and $\alpha=0$ independent of the dimension. In this 
case hyperscaling is fulfilled only at $d=4$.

The Josephson's relation is used experimentally to determine the value 
of $\nu$. Today it is possible to perform very accurate direct 
measurements of the penetration depth. The critical region probed 
is such that the gauge field fluctuations are unimportant and the 
critical fluctuations are those of the order parameter field 
and this means 
that $\eta_{A}^{*}=0$. Then, from Eq. (\ref{mA}), we obtain that near 
the phase transition $m_{A}^2\sim m^{d-2}\sim t^{\nu(d-2)}$, that is, 
$\rho_{s}\sim\lambda^{-2}\sim t^{\nu(d-2)}$. Thus, in such critical 
regime 
\begin{equation}
\label{nu'}
\nu'=\frac{\nu(d-2)}{2}.
\end{equation} 
Experiments performed in 
YBCO crystals give that $\nu'=0.33\pm 0.01$ \cite{Kamal,Kamal1}. Using 
Eq. (\ref{nu'}) for $d=3$ we obtain $\nu\approx 2/3$, consistent with 
the $3D$ $XY$ behavior, that is, a $^4$He like behavior. 

The situation seems to be however different for YBCO thin films. A 
recent measurement by Paget {\it et al.} \cite{Paget} performed in 
YBCO thin films gives $\nu'=1/2$, that is, a mean field like behavior. 
This result has been obtained to within  
$0.2-0.5$ K of $T_{c}$, indicating in this 
way a much smaller critical region as compared to the bulk YBCO 
\cite{Salamon,Kamal,Kamal1}. However, 
this mean-field like behavior could be interpreted 
as a non-mean field behavior in the following way. 
If we insist in using (\ref{nu'}) to 
evaluate $\nu$ we obtain     

\begin{equation}
\label{nu1}
\nu=\frac{1}{d-2},
\end{equation}
and, assuming that hyperscaling holds, 

\begin{equation}
\label{alpha}
\alpha=\frac{d-4}{d-2},
\end{equation}
The above exponents are not classical and we have $\nu=1$ and 
$\alpha=-1$ for $d=3$. 
Note the similarity between the 
present critical regime with the $O(n)$ model for large $n$ and 
$2<d\leq 4$. The above exponents are just the exact exponents for the 
$O(n)$ model at large $n$ \cite{ZJ}. The exponent $\nu$ as given in 
(\ref{nu1}) is also obtained in a $O(n)$ non-linear $\sigma$-model in 
$d=2+\epsilon$ dimensions and $n>2$ \cite{ZJ}. Thus, it is possible to 
interpret 
the experiments of Paget {\it et al.} as correponding to a non-classical 
situation characteristic of superconducting thin films. There are, however, 
some possible handicaps in this point of view. For instance, 
it is not expected 
a so expressive change in $\alpha$ for thin films relative to the 
bulk material. Thus, if $\alpha$ remains close to zero, the critical 
behavior probed in ref. \cite{Paget} corresponds in fact to mean-field. 
Another important point is the possibility of dimensional crossover 
behavior due to finite size effects arising from the thikness of the film. 
This situation has been extensively studied in superfluid 
$^4$He films \cite{Barber,Janke,Crowell}. For superconducting YBCO 
thin films the situation is unclear because in the optimally doped 
case the coupling between the $CuO$ planes is strong and the 2D 
fluctuations are probably dominated by the 3D fluctuations even for a small 
film thikness. However, in the underdoped cuprates the coupling 
between the $CuO$ planes is much weaker and we can expect a strong 
influence of 2D fluctuations for sufficiently small film thikness.      
 
To conclude this section, 
let us discuss the scaling behavior of the order parameter. This is a 
controversial matter both from the theoretical and experimental point of 
view. The theoretical controversy has its origin in the scaling behavior 
of the correlation function 
$W^{(2)}_{0}({\bf x},{\bf y})=<\phi_{0}({\bf x})\phi_{0}^{*}({\bf y})>$ 
at large 
distances, $|{\bf x}-{\bf y}|\to\infty$. For $|{\bf p}|\to 0$ and 
$m=0$ its Fourier 
transform $\tilde{W}^{(2)}_{0}$ behaves as 
$\tilde{W}^{(2)}_{0}({\bf p})\sim|{\bf p}|^{\eta-2}$. Most calculations 
gives a value of $\eta$ in the range $-1<\eta<0$ 
\cite{Herbut,deCalan,Berg,HLM,Hikami,Radz,Folk}. This does not contradicts 
the scaling relation $\beta=\nu(1+\eta)/2$ (for $d=3$) because 
$-1<\eta<0$ imply $\beta>0$ as it should be (note that $\nu$ must be 
positive). However, a negative value of $\eta$ is pathological in many 
respects. For instance, it has been pointed out by Kiometzis and Schakel 
\cite{KiometzisIII} that $\eta<0$ violates the positivity of the spectral 
weight in the K\"allen-Lehmann spectral representation of $W^{(2)}_{0}$ 
\cite{GJ}. In fact, this representation of $W^{(2)}_{0}$ implies that 
$0<Z_{\phi}<1$. Since near the critical point $Z_{\phi}\sim m^{\eta}$, we 
have necessarily that $\eta>0$. 
Moreover, from the $|{\bf p}|\to 0$ behavior of 
$\tilde{W}^{(2)}_{0}({\bf p})$ 
we see that $\eta<0$ makes the low momentum 
behavior worse than before renormalization \cite{Magnen}. This is in 
contradiction with the infrared stability of the fixed point. Another 
important point is that $\eta$ is the fixed point value of $\eta_{\phi}$, 
a quantity which is gauge dependent \cite{Nogueira}. Thus, we can question  
the physical meaning of the $\eta$ exponent and, as a consequence, 
of the order parameter itself. This situation is unconfortable 
because in the Gorkov derivation of the GL model from the BCS 
theory \cite{Gorkov}, $\phi$ is defined as being proportional to the gap 
function. In the microscopic theory the gap function has a precise 
physical meaning since it is responsible for the generation of a gap in 
the spectrum. The expectation value of $\phi$ is thus proportional to the 
gap near the critical temperature. The gauge dependence of the 
superconducting order parameter has been discussed recently in 
\cite{Nogueira} and it has been shown that 
$\partial\eta_{\phi}/\partial\alpha\to 0$ as the critical point is 
approached ($\alpha$ is the gauge fixing parameter), the only 
gauge contribution 
left to $\eta$ corresponding to the Coulomb gauge, $\alpha=0$. This result 
agrees with an early analysis by Kennedy and King 
\cite{Kennedy}, although these authors used a different, but closely related, 
definition of order parameter.  
Thus, the gauge dependence is 
actually not an issue and the point to be solved 
in the approximations is the negativeness of 
$\eta$.  

\section{Scaling and duality}

In this section we will study the scaling behavior of superconductors in a 
continuum dual Ginzburg-Landau (dGL) model. The dGL model has been proposed 
using plausible arguments on the dynamics of a vortex gas 
\cite{Kiometzis,Kovner,KiometzisI} and is assumed to be the continuum version 
of the dual GL model in the lattice \cite{Peskin,Dasgupta}.  
Lattice duality studies in abelian gauge models 
\cite{Dasgupta,Peskin,Bart}  
helped condensed matter theorists to obtain important conclusions concerning 
the superconducting phase transition. In particular, it has been used to   
establish that a second order phase transition should take place at least  
in the type II regime \cite{Dasgupta}.

\subsection{Duality in the lattice GL model}
 
For the sake of clarity, we set up 
here the arguments given in several papers 
\cite{Peskin,Dasgupta,Bart,KiometzisI,HerbutI}. 

A lattice version of the GL model has a partition function given by 

\begin{equation}
\label{lattice1}
Z(\beta,e)=\int_{-\pi}^{\pi}\left[\prod_{i}\frac{d\theta_{i}}{2\pi}\right]
\int_{-\infty}^{\infty}\left[\prod_{i,\mu}dA_{i\mu}\right]
\exp\left[\beta\sum_{i\mu}\cos(\Delta_{\mu}\theta_{i}-eA_{i\mu})
-\frac{1}{2}\sum_{i}({\bf \Delta}\times{\bf A}_{i})^2\right],
\end{equation}
where $\Delta_{\mu}\theta_{i}=\theta_{i+\hat{\mu}}-\theta_{i}$, that is, 
$\Delta_{\mu}$ is a lattice derivative. Note that in the above partition 
function the integral over $A_{i\mu}$ is over the interval 
$(-\infty,\infty)$, meaning that the gauge group is $\RR$, a non-compact 
gauge group. 

In the Villain approximation \cite{Villain} we can rewrite (\ref{lattice1}) 
as

\begin{equation}
\label{lattice2}
Z(\beta,e)=
\int_{-\pi}^{\pi}\left[\prod_{i}\frac{d\theta_{i}}{2\pi}\right]
\int_{-\infty}^{\infty}\left[\prod_{i,\mu}dA_{i\mu}\right]\sum_{m_{i\mu}}
\exp\left[-\frac{\beta}{2}
\sum_{i\mu}(\Delta_{\mu}\theta_{i}-eA_{i\mu}-2\pi m_{i\mu})^2
-\frac{1}{2}\sum_{i}({\bf \Delta}\times{\bf A}_{i})^2\right].
\end{equation}              
Let us introduce an auxiliary field $b_{i\mu}$ such that 

\begin{equation}
\exp\left[-\frac{\beta}{2}(\Delta_{\mu}\theta_{i}-eA_{i\mu}-2\pi m_{i\mu})^2
\right]\propto\int_{-\infty}^{\infty}db_{i\mu}
\exp\left[-\frac{1}{2\beta}b_{i\mu}^2+i(
\Delta_{\mu}\theta_{i}-eA_{i\mu}-2\pi m_{i\mu})b_{i\mu}\right].
\end{equation}
In the following we will neglect factors of proportionality which are 
generally smooth functions of $\beta$. Thus, many equations will be 
assumed up to proportionality factors. By performing the $\theta$ integrals 
we have

\begin{eqnarray}
Z(\beta,e)&=&\int_{-\infty}^{\infty}\left[\prod_{i,\mu}dA_{i\mu}\right]
\int_{-\infty}^{\infty}\left[\prod_{i,\mu}db_{i\mu}\right]
\delta({\bf \Delta}\cdot{\bf b}_{i})\sum_{m_{i\mu}}\exp\left[
\sum_{i\mu}\left(-\frac{1}{2\beta}b_{i\mu}^2+ieA_{i\mu}b_{i\mu}
-2\pi im_{i\mu}b_{i\mu}\right)\right.\nonumber\\
&-&\left.\frac{1}{2}\sum_{i}({\bf \Delta}\times{\bf A}_{i})^2\right].
\end{eqnarray}
Applying the Poisson formula we obtain 

\begin{equation}
\label{dual1}
Z(\beta,e)=\int_{-\infty}^{\infty}\left[\prod_{i,\mu}dA_{i\mu}\right]
\sum_{n_{i\mu}}\delta_{{\bf \Delta}\cdot{\bf n}_{i},0}
\exp\left[\sum_{i\mu}\left(-\frac{1}{2\beta}n_{i\mu}^2+ieA_{i\mu}n_{i\mu}
\right)-\frac{1}{2}\sum_{i}({\bf \Delta}\times{\bf A}_{i})^2\right].
\end{equation}
The constraint ${\bf \Delta}\cdot{\bf n}_{i}=0$ in (\ref{dual1}) implies 
that the links $n_{i\mu}$ form closed loops. These closed loop are 
interpreted as magnetic vortices \cite{KiometzisI,Peskin}. It can be shown 
that the $XY$ model partition function in the Villain approximation 
(Eq. (\ref{lattice2}) with $e=0$) can be cast in the form \cite{Peskin} 

\begin{equation}
\label{Zxy}
Z_{XY}(\beta)=\int_{-\infty}^{\infty}\left[\prod_{i\mu}da_{i\mu}\right]
\sum_{M_{i\mu}}\delta_{{\bf \Delta}\cdot{\bf M}_{i},0}\exp\left[
\sum_{i}-\frac{1}{2\beta}({\bf \Delta}\times{\bf a}_{i})^2+2\pi i
\sum_{i\mu}a_{i\mu}M_{i\mu}\right], 
\end{equation}
and we have $Z(\infty,e)=Z_{XY}(e^2/(2\pi)^2)$. $\beta<\infty$ corresponds 
to add a chemical potential for the loop variables in the 
dual $XY$ model. In 
this sense we can think of the lattice GL model in (\ref{lattice2}) as 
a generalized dual $XY$ model. Precisely, we have 
$Z_{XY}(\beta)=Z'_{XY}(\beta,0)$ where $Z'_{XY}(\beta,K)$ is the 
partition function of a generalized dual $XY$ model given by

\begin{equation}
\label{Zxy'}
Z'_{XY}(\beta,K)=\int_{-\infty}^{\infty}\left[\prod_{i\mu}da_{i\mu}\right]
\sum_{M_{i\mu}}\delta_{{\bf \Delta}\cdot{\bf M}_{i},0}\exp\left[
\sum_{i}-\frac{1}{2\beta}({\bf \Delta}\times{\bf a}_{i})^2+2\pi i
\sum_{i\mu}a_{i\mu}M_{i\mu}-K\sum_{i\mu}M_{i\mu}^2\right]. 
\end{equation}
It is possible to study the phase transition of the above dual model 
by looking the phase diagram in the $\beta-K$ plane \cite{Dasgupta}. In 
this phase diagram the point $(\beta_{c},0)$ corresponds to the $XY$ 
critical point, $\beta_{c}$ being the inverse critical temperature. From 
Eq. (\ref{lattice2}) with $e=0$ we obtain after integration of the 
$\theta$ variables 

\begin{equation}
\label{Zxydual}
Z_{XY}(\beta)=Z(\beta,0)=\sum_{m_{i\mu}}\delta_
{{\bf \Delta}\cdot{\bf m}_{i},0}\exp\left(-\frac{1}{2\beta}\sum_{i\mu}
m_{i\mu}^2\right).
\end{equation}
On the other hand, integration of the $a_{i\mu}$ in Eq. (\ref{Zxy'}) 
yields

\begin{equation}
\label{Zxy'1}
Z'_{XY}(\beta,K)=\sum_{M_{i\mu}}\delta_{{\bf \Delta}\cdot{\bf M}_{i},0}\exp\left[-2\pi^2\beta\sum_{i,j,\mu}M_{i\mu}G(|{\bf x}_{i}-
{\bf x}_{j}|)M_{j\mu}-K\sum_{i\mu}M_{i\mu}^2\right],
\end{equation} 
where the lattice Green function $G$ behaves like 
$|{\bf x}_{i}-{\bf x}_{j}|^{-1}$ for $|{\bf x}_{i}-{\bf x}_{j}|\to\infty$. 
From Eqs. (\ref{Zxydual}) and (\ref{Zxy'1}) we have that the point 
$(0,1/2\beta_{c})$ in the $\beta-K$ plane corresponds to the so called 
``inverted'' $XY$ transition. By doing the $A_{i\mu}$ integration in 
(\ref{dual1}) we obtain 

\begin{equation}
Z(\beta,e)=Z'_{XY}\left(\frac{e^2}{4\pi^2},\frac{1}{2\beta}\right), 
\end{equation}
which means that the lattice GL model will undergo an ``inverted'' 
$XY$ transition for $0<e^2<e^2_{c}=4\pi^2\beta_{c}$ \cite{Dasgupta}. 

Let us work out further Eq. (\ref{dual1}). Since 
${\bf \Delta}\cdot{\bf n}_{i}=0$ we can put 
${\bf n}_{i}={\bf \Delta}\times{\bf l}_{i}$. By using the Poisson formula 
to introduce a continuum field $h_{i\mu}$ and integrating out the 
$A_{i\mu}$'s, we obtain 

\begin{equation}
\label{dual2}
Z(\beta,e)=\int_{-\infty}^{\infty}\left[\prod_{i\mu}dh_{i\mu}\right]
\sum_{m_{i\mu}}\delta_{{\bf \Delta}\cdot{\bf m}_{i},0}\exp\left\{
\sum_{i}\left[-\frac{1}{2\beta}({\bf \Delta}\times{\bf h}_{i})^2
-\frac{e^2}{2}{\bf h}_{i}^2\right]+2\pi i\sum_{i\mu}m_{i\mu}h_{i\mu}\right\}. 
\end{equation}
Note that putting $e=0$ in (\ref{dual2}) we obtain a partition function 
identical to (\ref{Zxy}), as it should be. 
Reintroducing the variable $\theta$ through

\begin{equation}
\label{delta}
\delta_{{\bf \Delta}\cdot{\bf m}_{i},0}=\int_{-\pi}^{\pi}
\frac{d\theta_{i}}{2\pi}
e^{i\theta_{i}({\bf \Delta}\cdot{\bf m}_{i})},\nonumber
\end{equation}
and using the Poisson formula to convert the integral over $h_{i\mu}$ in a 
sum over $n_{i\mu}$, we obtain

\begin{equation}
\label{dual}
Z(\beta,e)=
\int_{-\pi}^{\pi}\left[\prod_{i}\frac{d\theta_{i}}{2\pi}\right]
\sum_{n_{i\mu}}\exp\left[
-\frac{e^2}{8\pi^2}\sum_{i\mu}(\Delta_{\mu}\theta_{i}-2\pi n_{i\mu})^2
-\frac{\beta}{2}\sum_{i}({\bf \Delta}\times{\bf n}_{i})^2\right].
\end{equation}
The partition function (\ref{dual}) is the dual representation of 
(\ref{lattice2}) used in the numerical simulations in ref. \cite{Dasgupta}. 
Note that the dual representation (\ref{dual}) has gauge group 
$\ZZ$. We observe that the $Z$ given in (\ref{lattice2}) with $e=0$ 
should be the same, up to proportionality constants, as the $Z$ given 
in (\ref{dual}) with $\beta=0$ provided we put $\beta=e^2/4\pi^2$ in 
(\ref{lattice2}). It is clear that the temperature 
$\beta^{-1}$ plays the role of a 
dual charge satisfying the Dirac condition $e\beta^{-1}=2\pi$. 

By performing the $h_{i\mu}$ integration in (\ref{dual2}) we obtain an 
equation analogous to (\ref{Zxy'1}) with $K=0$. The difference is that 
the Green function $G$ is replaced by a massive Green function 
$\tilde{G}(|{\bf x}_{i}-{\bf x}_{j}|)$ which behaves like 
$e^{-\beta e^2|{\bf x}_{i}-{\bf x}_{j}|}/|{\bf x}_{i}-{\bf x}_{j}|$ for 
$|{\bf x}_{i}-{\bf x}_{j}|$ large. Thus, we can write

\begin{equation}
Z(\beta,e)=Z'_{XY}\left(\frac{e^2}{4\pi^2},\frac{1}{2\beta}\right)
=\sum_{m_{i\mu}}\delta_{{\bf \Delta}\cdot{\bf m}_{i},0}\exp
\left[-2\pi^2\beta\sum_{i,j,\mu}m_{i\mu}
\tilde{G}(|{\bf x}_{i}-{\bf x}_{j}|)m_{j\mu}\right].
\end{equation}
Therefore, adding a chemical potential to the loop variables in the 
dual $XY$ model is equivalent to replace the massless Green function $G$ 
by a massive one in a dual $XY$ model with $K=0$. In this duality map we have 
represented the partition function of the lattice GL model in such a way 
that the loop variables have zero chemical potential. This amounts 
in replacing the massless Green function by a massive one. This means that 
if we consider a scaling (continuum) limit of the dual model, it is 
necessary to consider a massive gauge field ${\bf h}({\bf x})$. 
This gauge field should satisfy the constraint 
$\nabla\cdot{\bf h}=0$ and should be coupled minimally to a field 
$\psi$ such that $|\psi|^2$ represents the density of magnetic vortices.

\subsection{The continuum dual model}      

The continuum version of lattice duality, the dual Ginzburg-Landau model
(dGL) has been proposed on the basis of plausible 
arguments concerning the dynamics of a vortex gas 
\cite{Kiometzis,Kovner,KiometzisI}.   
Attempts have been made to justify it as the 
continuum limit of the lattice dual model 
\cite{KiometzisI,HerbutI} but it does not 
exist to date a rigorous mathematical construction of 
the continuum limit of the lattice dual model. For instance, a possible 
way towards motivating the dGL model from the lattice dual model has 
been proposed by Herbut \cite{HerbutI}. Instead combining (\ref{dual2}) 
and (\ref{delta}) to obtain (\ref{dual}), we can follow Peskin 
\cite{Peskin} and insert in (\ref{dual2})

\begin{equation}
1=\lim_{t\to 0}\exp\left(-\frac{t}{2}\sum_{i\mu}m_{i\mu}^2\right).
\end{equation}
By using (\ref{delta}) and the identity 

\begin{equation}
\label{ident}
\sum_{m=-\infty}^{\infty}e^{-\frac{t}{2}m^2+ixm}=
\sqrt{\frac{2\pi}{t}}\sum_{M=-\infty}^{\infty}e^{-\frac{1}{2t}
(x-2\pi M)^2}, 
\end{equation}
we obtain 

\begin{eqnarray}
\label{Herbut}
Z(\beta,e)&=&\lim_{t\to 0}\left(\frac{2\pi}{t}\right)^{N/2}
\int_{-\infty}^{\infty}\left[\prod_{i\mu}dh_{i\mu}\right]
\int_{-\pi}^{\pi}\left[\prod_{i}\frac{d\theta_{i}}{2\pi}\right]
\sum_{m_{i\mu}}\exp\left[-\sum_{i\mu}\frac{1}{2t}(\Delta_{\mu}
\theta_{i}-2\pi h_{i\mu}-2\pi M_{i\mu})^2\right.\nonumber\\
&-&\left.-\frac{1}{2\beta}({\bf \Delta}\times{\bf h}_{i})^2
-\frac{e^2}{2}{\bf h}_{i}^2\right].
\end{eqnarray}
The limit $t\to 0$ generates delta functions in the integrand which, 
when replaced by the integral representation and applying the 
Poisson formula allows us to recover (\ref{dual2}). It has been proposed 
in \cite{HerbutI} that by leaving $t$ finite Eq. (\ref{Herbut}) is 
analogous to (\ref{lattice2}), with the difference that in (\ref{Herbut}) 
the lattice gauge field is massive. Moreover, in virtue of (\ref{ident}) 
$t$ can be interpreted as a chemical potential for the loop variables. 
Thus, by keeping $t$ small but finite we can write the following 
continuum limit of (\ref{Herbut}) \cite{Kiometzis,HerbutI,Kovner,KiometzisI} 
in terms of bare quantities:  

\begin{equation}
\label{Sd}
S=\int d^{3}x\left[\frac{1}{2}
(\nabla\times{\bf h}_{0})^2
+\frac{m_{h,0}^2}{2}
{\bf h}_{0}^2
+|(\nabla-im_{h,0}q_{0}{\bf h}_{0})\psi_{0}|^2+
\mu_{0}^2|\psi|^2+\frac{v_{0}}{2}|\psi_{0}|^4\right].
\end{equation}
In (\ref{Sd}), the bare {\it dual} charge, $q_{0}$, is related to the charge 
$e_{0}$ by the Dirac condition $q_{0}e_{0}=2\pi$. The 
field $\psi$ is usually called a {\it disorder parameter}, as opposed to 
the order parameter in the GL model.
We can of course criticize the motivation of (\ref{Sd}) from (\ref{Herbut}) 
because it is based on the hypothesis of a finite, though small $t$ in 
(\ref{Herbut}). In fact, (\ref{Herbut}) represents a lattice GL model 
{\it only if} $t\to 0$. A small nonzero $t$ constitutes an approximation 
and the continuum limit above should be regarded as an approximate 
continuum dual model. This should be contrasted with the {\it exact} 
duality map we have obtained in the lattice (up to a Villain approximation). 
The dGL model given by (\ref{Sd}) has been also motivated by arguing 
directly in the continuum but using the London limit of the GL model, that is, 
by assuming that $\phi({\bf x})=\bar{\phi}e^{i\theta}$ with 
$\bar{\phi}$ constant \cite{KiometzisI}. 
All of this is approximate and we cannot really say 
that the dual map in the continuum is exact. The construction of 
continuum limits is a highly non-trivial matter, even for a simple scalar 
model \cite{GJ}. 
 
Although the dGL model gives a respectable value for the correlation lenght 
exponent \cite{Kiometzis,HerbutI}, it does not give $\nu'=\nu$ 
when $m_{h,0}$ is identified with the bare photon mass $m_{A,0}$ of the 
GL model \cite{Kiometzis}. 
Depending on the choice of scaling (to be precised later), we 
can find a different value of $\nu'$, as will be shown in the next 
paragraphs. We will consider a family of scalings $\Sigma_{\zeta}$ where 
$\zeta\geq 0$ is a parameter. For each $\zeta$ we 
find a different phase transition regime that must correspond to a possible 
critical behavior 
in the GL model. By considering this family of scalings, we will 
show that it is not possible to describe an ``inverted'' $XY$ behavior with 
$m_{h,0}=m_{A,0}$ and, at the same time, $m_{h}=m_{A}$, $m_{h}$  
being the renormalized counterpart of $m_{h,0}$.

We define 
the renormalized fields ${\bf h}=Z_{h}^{-1/2}{\bf h}_{0}$ and 
$\psi=Z_{\psi}^{-1/2}\psi_{0}$. From the Ward identities we obtain that the 
term $m_{h,0}^2{\bf h}_{0}^2/2$ 
does not renormalize, implying in this way that 
$m_{h}^2=Z_{h}m_{h,0}^2$.  
The renormalization of the remaining 
parameters follows easily: $\mu^2=(Z^{(2)}_{\psi})^{-1}Z_{\psi}\mu_{0}^2$, 
$v=Z_{v}^{-1}Z_{\psi}^2 v_{0}$ and $q=q_{0}$. 
We note that the dual charge remains bare. Let us define the 
quantity $g_{0}^2=m_{h,0}^2 q_{0}^2$ which renormalizes as 
$g^2=Z_{h}g_{0}^2$. 
The dimensionless couplings relevant to the problem are 
$\hat{g}^2=g^2/\mu$ and $\hat{v}=v/\mu$. We will perform our RG analysis 
by introducing  
a scaling hypothesis for $m_{h,0}$.  
We assume that $m_{h,0}^2\propto t^{\zeta}$, where 
$\zeta\geq 0$. This scaling hypothesis will define under a RG a 
family of scalings $\Sigma_{\zeta}$. It is useful to define the ratio 
$\kappa_{d}=\mu/m_{h}$, analogous to the Ginzburg parameter $\kappa$ in the 
GL model. For arbitrary $\zeta\geq 0$ we obtain the flow equations: 

\begin{equation}
\label{kd}
\mu\frac{\partial\kappa_{d}^2}{\partial\mu}=[2-\eta_{h}-\zeta(2+
\eta_{\psi}^{(2)}-\eta_{\psi})]\kappa_{d}^2,
\end{equation}         
 
\begin{equation}
\label{mh}
\mu\frac{\partial m_{h}^2}{\partial\mu}=
[\eta_{h}+\zeta(2+\eta_{\psi}^{(2)}-\eta_{\psi})]m_{h}^2,
\end{equation}

\begin{equation}
\label{g2}
\mu\frac{\partial\hat{g}^2}{\partial\mu}=
[\eta_{h}-1+\zeta(2+\eta_{\psi}^{(2)}-\eta_{\psi})]\hat{g}^2, 
\end{equation}
where the RG functions $\eta_{\psi}^{(2)}$, $\eta_{\psi}$ and 
$\eta_{h}$ are 
defined by 

\begin{equation}
\eta_{\psi}^{(2)}=\mu\frac{\partial\ln Z_{\psi}^{(2)}}{\partial\mu},
\end{equation}

\begin{equation}
\eta_{\psi}=\mu\frac{\partial\ln Z_{\psi}}{\partial\mu},
\end{equation}

\begin{equation}
\eta_{h}=\mu\frac{\partial\ln Z_{h}}{\partial\mu}.
\end{equation}

Near the phase transition 
$\mu=\xi^{-1}\sim t^{\nu}$. As $\zeta$ varies, the above equations 
describe different regimes of phase transitions in the case where an infrared 
stable fixed point exists. Let us analyse some relevant cases. 

\subsubsection{$\Sigma_{0}$ scaling}
For 
$\zeta=0$ we have that the infrared stable fixed point for Eq. (\ref{g2}) 
is a non-vanishing $\hat{g}^2_{*}$ satisfying 
$\eta_{h}^{*}\equiv\eta_{h}(\hat{g}^2_{*},\hat{v}_{*})=1$, 
$\beta_{\hat{v}}(\hat{g}^2_{*},\hat{v}_{*})=0$ with 
$\hat{v}_{*}$ being the fixed point value of the coupling $\hat{v}$. From 
Eq. (\ref{mh}) we obtain that near the phase transition 
$m_{h}\sim\mu^{1/2}$. By defining the exponent of $m_{h}$ through 
$m_{h}\sim t^{\nu_{h}}$, we obtain that $\nu_{h}=\nu/2$. This 
suggests that $\kappa_{d}\to 0$ as approaching the critical point. This is 
the case indeed, since from Eq. (\ref{kd}) we obtain that near the 
infrared stable fixed point $\kappa_{d}\sim\mu^{1/2}$. From this behavior of 
$\kappa_{d}$ we obtain that the fixed point value $\hat{v}_{*}$ 
corresponds to the value in a $XY$ model and $\nu=\nu_{XY}\approx 2/3$. 
The reason for this behavior comes from the fact that every power of the 
dual charge in $\beta_{\hat{v}}$ is multiplied by a function of $\kappa_{d}$ 
which tends to zero as the critical point is approached. The same type of 
behavior has been already encountered in the litterature 
\cite{HerbutI,Nogueira}. 
If we identify 
$m_{h}$ with the renormalized photon mass $m_{A}$ 
we obtain $\nu'=\nu_{XY}/2$, the 
half of the value expected for the superconducting transition in the 
``inverted'' $XY$ universality class \cite{Olsson}. However, this result 
corresponds to the universality class of the crossover regime governed 
by the {\it neutral} 
$XY$ fixed point \cite{Fisher}, which is the only one we have 
experimental access \cite{Salamon,Kamal}.    

\subsubsection{$\Sigma_{1}$ scaling}
We consider now the scaling $\Sigma_{1}$ corresponding to 
$\zeta=1$. In this case we obtain that the infrared stable fixed 
point corresponds to $\hat{g}^2_{*}=0$. The present situation matchs  
with the one encountered in \cite{Kiometzis} where $m_{h,0}$ is 
identified with the bare photon mass $m_{A,0}$. We have still that 
$\nu=\nu_{XY}$. However, if $m_{h}=m_{A}$ we obtain that 
$\nu_{h}=\nu'=1/2$ and the penetration depth exponent is classical. 
As we have already discussed in section II, the value $\nu'=1/2$ has 
been measured by Lin {\it et al.} \cite{Lin} and in a recent paper 
by Paget {\it et al.} \cite{Paget}.  

\subsubsection{$\Sigma_{\zeta}$ scaling with $\zeta\approx 4/3$}
Clearly the scalings $\Sigma_{0}$ and $\Sigma_{1}$ of the dGL 
model are not in the 
``inverted'' $XY$ universality class but belong to 
different crossover regimes in the GL model. In order to have an  
``inverted'' $XY$ behavior we need to obtain $\nu_{h}=\nu'=\nu=\nu_{XY}$, 
after identifying $m_{h}$ with the renormalized photon mass in the GL model. 
It is easy to check that this is in fact the case for 
$\zeta\approx 4/3$, but in this case $m_{h,0}$ clearly does not 
corresponds to the bare photon mass. Rather, $m_{h,0}$ behaves like the 
renormalized photon mass.  
For $\zeta\approx 4/3$, the infrared stable fixed point 
corresponds to a neutral dual charge just as in the scaling $\Sigma_{1}$  
and from Eq. (\ref{mh}) we obtain $m_{h}\sim\mu$ near the fixed 
point. Note that in contrast with the other two cases, now we have that 
$\kappa_{d}$ approaches a non-vanishing fixed 
point value in the infrared.                     

\section{Concluding remarks}

We have discussed in this paper some relevant aspects of the scaling 
behavior of superconductors in a zero magnetic field Meissner phase. 
We have obtained in the first part of the paper  
the scaling relation $\nu'=\nu$ as a consequence of 
an exact flow equation for the photon mass. The interesting point of 
our derivation is that since no use has been done of the Josephson 
relation, this relation is obtained directly in the form 
$\rho_{s}\sim t^{\nu}$. We argued that this implies the 
validity of the hyperscaling relation 
for a charged superfluid since the Josephson relation 
can be derived independently in the form $\rho_{s}\sim t^{2\beta-\nu\eta}$. 
In this part some emphasis has been given to the 
relation between theory and experiment, particularly in what concerns 
the usefulness of the Josephson relation in determining the exponent 
$\nu$. While this is not new, it helped us to propose an 
alternative interpretation 
of a recent experiment performed by Paget {\it et al.} \cite{Paget} 
in YBCO thin films. We 
proposed that the correlation lenght and specific heat exponents 
are possibly non-classical and given respectively by $\nu=1$ and 
$\alpha=-1$. In order to confirm this it is necessary to measure 
$\alpha$ directly in thin films, which is a very difficult task. We are 
aware however of the little probability of changing so dramatically the 
value of $\alpha$ with respect to the bulk sample value. Anyway, such 
a possibility cannot be completely discarded and we hope that some 
experimental effort could be made in this sense.     

In the second part of the paper we tried to elucidate the scaling 
behavior of the continuum dual Ginzburg-Landau model. 
This has been done by the introduction of a scaling hypothesis on the 
bare magnetic induction mass. Our arguments in this part of the paper 
show that the continuum dual model still deserves more reflexion. 
We think that the dual map in the continuum is not as complete as it is in  
the lattice. We emphasize that there is no rigorous argument in favor 
of the usual continuum model (\ref{Sd}).      
A future theoretical perspective is to study different scalings in 
another possible choice for a 
continuum dual model. It consists of a continuum $XY$ model where  
a {\it perturbatively} non-renormalizable interaction of the form 
$(\psi^{\dag}\nabla\psi-\psi\nabla\psi^{\dag})^2$ \cite{Kovner,KovnerI}  
is added. It is possible that 
such an interaction can be non-perturbatively renormalizable, like the 
situation encountered in the Gross-Neveu model in $d=3$ \cite{Ros} where 
the non-perturbative renormalizability has been rigourously 
demonstrated \cite{deCalanI}.   

\acknowledgments

The authors would like to acknowledge J. Magnen for discussions. FSN 
would like to thank Prof. T. R. Lemberger for sending 
some additional comments concerning the work of his group.

\end{document}